\documentclass[%
 reprint,
 amsmath,amssymb,
 aps,
 prl
]{revtex4-1}

\usepackage{graphicx}% Include figure files
\usepackage{dcolumn}% Align table columns on decimal point
\usepackage{bm}% bold math
\begin{document}

%\preprint{APS/123-QED}

\title{Line Tension Reshapes Nucleation at Surface Edges: \\ A Generalized Theory for Nanopore Activation}
\author{Yanchen Wu}
\author{Martin Z. Bazant}
\author{Allan S. Myerson}
\author{Richard D. Braatz}
\email{braatz@mit.edu}
\affiliation{Department of Chemical Engineering, Massachusetts Institute of Technology, 77 Massachusetts Avenue, Cambridge, MA 02139, USA}

%RDB: what do you think about inviting bernhart l. trout to review the manuscript?

%\affiliation {Department of Chemical Engineering, Massachusetts Institute of Technology, Cambridge, MA 02139}
%\affiliation
%{Center for Computational Science and Engineering, Massachusetts Institute of Technology, Cambridge, MA 02139}

\date{\today}% It is always \today, today,

\begin{abstract}
Heterogeneous nucleation at surface edges is pervasive across nature and industry, yet the role of line tension, arising from asymmetric capillary interactions at geometric singularities, remains poorly understood. Herein we develop a generalized nucleation theory that explicitly incorporates line tension induced by edge pinning, thereby extending classical frameworks to account for nanoscale confinement and interfacial asymmetry. Through analytical treatment of droplet formation within geometrically defined nanopores, we derive a closed-form expression for the edge-pinned line tension as a function of Laplace pressure, pore geometry, and wettability. This formulation reveals that line tension can significantly reshape the nucleation energy landscape, introducing nontrivial dependencies on contact angle and pore morphology. Our results uncover a tunable, geometry-mediated mechanism for controlling nucleation barriers, offering predictive insight into phase transitions in confined environments and suggesting new strategies for design in applications ranging from nanofluidics to crystallization control.
%is an intensive surface property and it is not a function of length of the contact line.
\end{abstract}

\maketitle

Nucleation, the emergence of a new daughter phase from a metastable parent phase~\cite{gibbs1878equilibrium,debenedetti2024special}, is fundamental to countless natural and technological processes, including cloud and ice  
 formation~\cite{dusek2006size,wilson2015marine,campbell2018nucleation}, amyloid fibril aggregation in Alzheimer’s disease~\cite{sevigny2016antibody}, pharmaceutical crystallization~\cite{guthrie2018controlling}, and nanofabrication~\cite{wu2022nucleation}. %and concrete curing~\cite{scherer2012nucleation}.  
%It occurs when thermal fluctuations overcome energy barriers governed by interfacial energies. 
For over a century, classical nucleation theory (CNT) and its heterogeneous extension (hetCNT) have provided the foundational framework for describing nucleation in bulk and at idealized surfaces~\cite{piucco2008study,fletcher1958size,gomez2015phase,narhe2004nucleation,cacciuto2004onset,singha2015thermokinetics}. However, these models rely on oversimplified assumptions of smooth, regular geometries and neglect nanoscale features such as edges, crevices, and confinement, which are ubiquitous in real systems.
In particular, the existence of line tension further complicates the problem and its
role in wetting and nucleation
remains poorly understood~\cite{gretz1966line,navascues1981line,amirfazli2004status,schimmele2007conceptual,wang2024wetting}.

Despite extensive efforts to infer line tension values from contact angle measurements~\cite{berg2010impact}, particle detachment experiments~\cite{kolarov1968contact}, and nucleation rate analyses~\cite{toshev1988line,hienola2007estimation}, its magnitude and even sign remain controversial~\cite{law1994theory,tan2025contact,moller2024crystal,buchanan2024kernel}.
Line tension plays a significant role at the nanoscale, influencing wetting~\cite{raspal2012nanoporous}, nucleation~\cite{auer2003line,tinti2017intrusion}, boiling~\cite{giacomello2023keeps}, and drying~\cite{guillemot2012activated} in confined geometries. Nanopores, in particular, exhibit a variety of confinement-induced phenomena. For instance, vapor nucleation within nanopores governs biological gating and ion transport~\cite{beckstein2003liquid,walther2013barriers,paulo2023hydrophobically}. 
Capillary condensation, freezing transition, and intrusion-extrusion processes  in confined pore geometries are strongly influenced by molecular layering~\cite{bazant2012theory}, electrostatic interactions~\cite{zhou2020freezing}, network connectivity~\cite{pinson2018inferring}, and contact angle hysteresis~\cite{pinson2018inferring}, all of which can modulate nucleation pathways beyond classical expectations. The long-term stability of surface nanobubbles—central to hydrophobic interactions in biomolecular systems—has been linked to line tension and contact line pinning on nanoscale surface features~\cite{liu2013nanobubble,lohse2015surface,sun2016stability,tan2021stability}.
Nevertheless, existing models such as interface displacement model often introduce line tension as a phenomenological correction to classical models~\cite{indekeu1992line,bauer1999quantitative,amirfazli2004status}, decoupled from the actual geometry constrain of the substrate and contact line pining, which may contribute to the long-standing ambiguity in its magnitude and sign. 
%This long-standing theoretical gap becomes especially critical in confined geometries 
%, which underpins key phenomena in both fundamental and applied science. 
\begin{figure}[b!]
\centering
\includegraphics[width=0.47\textwidth]{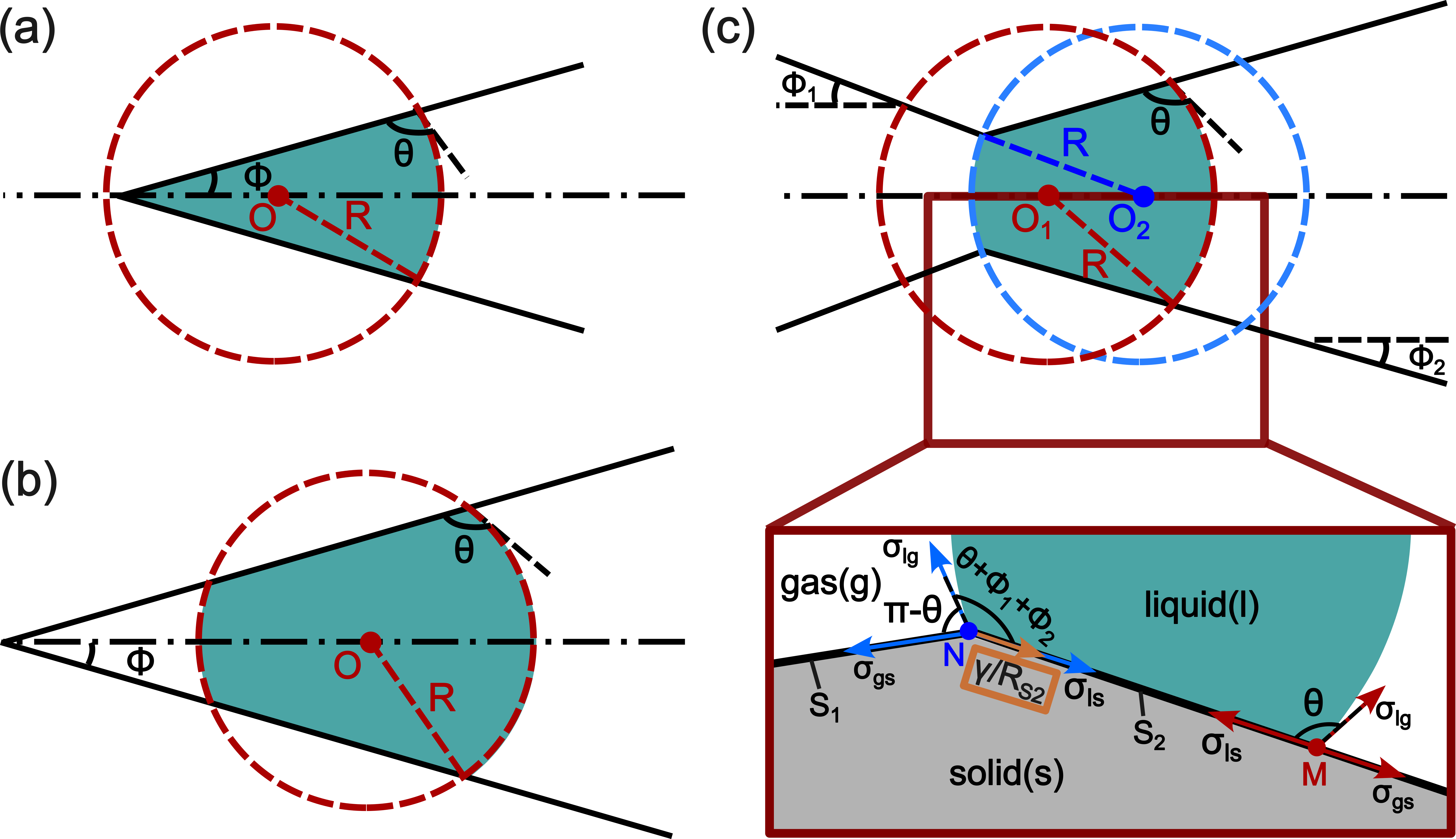}% Here is how to import EPS art
\caption{Schematic cross-sectional view of critical nuclei (green, with radius $R$) forming within pore structures with given pore opening angles 2$\phi$ (or 2$\phi_1$ and 2$\phi_2$) and contact angles $\theta$. (a) and (b) illustrate classical scenarios in smooth conical pores.  (c) depicts a generalized asymmetric pore geometry, where a critical droplet bridges the neck between two sides with distinct opening angles (left: 2$\phi_1$; right: 2$\phi_2$).  Red and blue dashed circles, centered at $O_1$ and $O_2$,
represent fitted curvature profiles on the right and left interfaces, respectively.  The lower part of (c) zooms in on the three-phase contact region, showing a force balance sketch at the triple contact line.  The points $M$ and $N$ denote the triple-phase contact locations, while $S_1$ and $S_2$ represent the inner pore surfaces. The surface edge at point $N$ gives rise to a line tension $\gamma$, and $R_{S2}$ is the curvature radius of the contact line at $N$ projected on surface $S_2$.}
\label{fig1:schematic-pore}
\end{figure}

Here we address this gap by developing a rigorous mathematical framework for heterogeneous nucleation that explicitly couples line tension to geometric confinement and contact line pinning.
Unlike classical treatments that attribute line tension solely to molecular interactions, our approach isolates the geometric or pinning-induced contribution. We demonstrate that such contributions arise intrinsically from geometric confinement, independent of intermolecular forces, and can significantly alter nucleation behavior in nanoscale environments.

Focusing on droplet formation in pore structures with defined opening angles, we derive an analytic expression for the line tension in terms of the Laplace pressure, geometric parameters, and wettability. This formulation reveals that the pinning induced line tension is not an intrinsic material constant, but a geometry-dependent quantity that fundamentally modifies the nucleation barrier. 
Our results show how pore geometry and surface wettability can be tuned to control nucleation pathways, regulating nucleation energy barriers on demand. It opens a path toward rational design of nucleation processes, offering a bridge between fundamental interfacial physics and real-world control over phase transitions in confined systems.

To capture how line tension emerges and reshapes the nucleation process in pore structures, we consider a quasi-equilibrium liquid droplet nucleating inside a pore with geometrically defined edges, e.g., conical or asymmetric constrictions as seen in Fig.~\ref{fig1:schematic-pore}. 
 The Gibbs free energy change related to the nucleation formation is 
\begin{equation}
\centering
\Delta G=\Delta F_H - V\Delta p,
\label{eq:gibbs}
\end{equation}
where $\Delta F_H$ is the variation in the Helmholtz free energy due to the formation of a cluster, which is formulated as
\begin{equation}
\centering
\Delta F_H=\sigma_{lg} A_{lg} + (\sigma_{ls}-\sigma_{gs})A_{ls}+ \gamma l,
\label{eq:helm}
\end{equation}
where $\sigma_{lg}$, $\sigma_{ls}$, and $\sigma_{gs}$ are interfacial tensions between liquid-gas, liquid-substrate, and gas-substrate, respectively. The parameters $A_{lg}$ and $A_{ls}$ denote liquid-gas and liquid-substrate interfacial areas, respectively. In the last term, $\gamma$ is the line tension and $l$ stands for the length of the triple contact line along the surface edge. 
Here, “line tension” is induced by an effective geometric energy correction $\Delta E_{\mathrm{geo}}$ arising from the coupling between Laplace pressure and edge-induced confinement, rather than a molecular-scale quantity. We define $\gamma = \Delta E_{\mathrm{geo}} / l$ to capture this scenario-dependent effect, offering insight into the long-standing ambiguity in its experimental determination.
%In the conventional case of smooth surfaces without pinning effects, the last term is typically neglected. However, for droplets pinned at surface edges, the line tension term becomes significant and can no longer be ignored. This will be discussed in detail in the following sections.

The second term in the right-hand side of Eq.~\eqref{eq:gibbs} is the volumetric free energy, in which $\Delta p$ is the pressure difference between the droplet and the surrounding supersaturated vapor. Its value is closely related to the supersaturation $S$ by
\begin{equation}
\centering
\Delta p=\frac{k_BT \ln S}{v_m},
\label{eq:delta-p}
\end{equation}
with $k_B$, $T$ and $v_m$ denoting the Boltzmann constant, temperature, and the molar volume, which are kept unchanged.

A nucleus will grow once the critical free energy barrier is overcome.
The critical parameters for heterogeneous nucleation can be determined by minimizing the Gibbs free energy change.
Thus, the critical nucleus corresponds to the condition $\partial \Delta G/\partial R=0$, i.e.,
\begin{equation}
\centering
\left. \frac{\partial \Delta F_H}{\partial R} \right|_{R=R_c} - \Delta p \left. \frac{\partial V}{\partial R} \right|_{R=R_c} = 0
\label{eq:critical1}
\end{equation}
where the critical nucleus radius $R_c$ satisfies the Young-Laplace equation $\Delta p=2\sigma_{lg}/R_c$.
Throughout this work, we assume a constant liquid–gas surface tension $\sigma_{lg}$, so that the critical nucleus radius $R_c$ depends solely on the pressure difference $\Delta p$ or equivalently, on the supersaturation $S$. Under the further assumption of constant supersaturation, $R_c$ is treated as a fixed value.
The wetting behavior at the substrate is typically characterized by the Young contact angle $\theta$, which is related to the interfacial tensions through Young’s equation $\cos\theta=(\sigma_{gs}-\sigma_{ls})/\sigma_{lg}$. 

Fig.~\ref{fig1:schematic-pore}(a) and (b) illustrate classical scenarios of critical nucleus formation in smooth conical pores, in the absence of surface edge pinning effects. Specifically, panel (a) corresponds to the case where the pore is completely filled by the droplet, while panel (b) depicts a droplet forming as a spherical cap, intersecting only a portion of the pore interior.
These classical configurations have been partially studied in previous works~\cite{willmott2011uptake,singha2015thermokinetics}, and a more detailed discussion is provided in the Supplemental Document.
They serve as a baseline for comparison with the generalized model developed in this work.

In contrast to classical approaches that consider smooth conical pores and neglect the contact line contribution, we explicitly resolve its geometry and force balance at the three-phase junction on the surface edge for a more generalized
pore geometry as shown in Fig.~\ref{fig1:schematic-pore}(c). In particular, we focus on pores with asymmetric opening angles, where sharp surface edges introduce geometric singularities that give rise to an additional capillary force that we interpret as line tension related force, accounting for the deviation of the apparent contact angle from the Young's contact angle on the droplet covered inner pore surface (denoted as $S_2$ in Fig.~\ref{fig1:schematic-pore}(c)) near the edge $N$.  
We assume that the droplet shows an apparent contact angle of  $\theta+\phi_1+\phi_2$ on surface $S_2$ whereas it forms an apparent angle of $\theta$ on surface $S_1$. At position $M$, far from the edge, Young’s equation applies and the local contact angle remains $\theta$.

Through detailed geometrical analysis, we derive expressions for the droplet volume $V=V(\theta, \phi_1, \phi_2, R)$ and the interfacial areas $A_{lg}$ and $A_{ls}$, as functions of the contact angle, geometric parameters, and droplet radius. By combining these expressions with Eq.~\eqref{eq:helm} and Eq.~\eqref{eq:critical1}, we obtain a closed-form expression for a dimensionless line tension, defined as $\gamma*=:\gamma/(\sigma_{lg} R_c)$. This expression serves as a predictive relation linking the critical radius of the nucleating droplet, the contact angle, and the local pore geometry (see Supplemental Document for the full derivation),
\begin{equation}
\centering
\gamma*=\frac{\gamma}{\sigma_{lg} R_c}=-\frac{\cos(\theta+\phi_1) }{\sin \phi_2}[\cos(\theta+\phi_1+\phi_2)-\cos\theta].
\label{eq:critical-linetension}
\end{equation}
This equation shares the same form as the classical modified Young's equation considering the line tension effect. The physical interpretation is that the line-tension-induced capillary force accounts for the deviation of the apparent contact angle $\theta + \phi_1 + \phi_2$ on surface $S_2$ from the intrinsic Young’s angle $\theta$. The prefactor $-\frac{R_c\cos(\theta+\phi_1) }{\sin \phi_2}$ corresponds exactly to the radius of curvature of the contact line at
$N$ projected onto surface $S_2$, denoted as $R_{s2}$ in Fig.~\ref{fig1:schematic-pore}(c).
The above geometric and force balance analysis reveals that this type of line tension emerges as a mechanical consequence of Laplace pressure under geometric confinement and pinning effects.

This insight, in turn, validates the inclusion of the line contribution in Eq.~\eqref{eq:helm}. It is noteworthy, however, that the length $l$ in Eq.~\eqref{eq:helm} specifically refers only to the contact line along the surface edge. Possible contributions from the contact line on smooth surfaces, where line tension can arise from intermolecular or long-range forces~\cite{joanny1986role}, have been neglected.

Building upon this theoretical framework of edge-pinned line tension, we introduce a geometric parameter $f(\theta, \phi_1, \phi_2)$ to characterize the ratio of the critical energy barrier in heterogeneous nucleation to that of the homogeneous nucleation, i.e.,
\begin{equation}
\centering
f(\theta, \phi_1, \phi_2)=\frac{ \Delta G_c}{\Delta G_{\text{hom},c}}=\sum_{i=1}^{3} f_i(\theta, \phi_1, \phi_2),
\label{eq:critical3}
\end{equation}
where $\Delta G_{\text{hom},c}$ and $f_i$ are
\begin{align}
\centering
\notag
 &\Delta G_{\text{hom},c}=\frac{4}{3}\pi R_c^2\sigma, \\
\notag
 &f_{1}= \frac{3\cos^2\theta_1}{2\sin\phi_2}[\cos(\theta +\phi_1 +\phi_2)-\cos\theta ] , \\
 \notag
 &f_{2}=\frac{3}{4} \Big[4-2\sin\theta_1-2\sin\theta_2-\frac{\cos\theta}{\sin\phi_2}(\cos^2\theta_2-\cos^2\theta_1)\Big], \\
 \notag
 &f_{3}=-\frac{1}{2} \Bigg[\sum_{i=1}^{2}(2+\sin\theta_i)(1-\sin\theta_i)^2+\frac{\cos^3\theta_2-\cos^3\theta_1}{\tan\phi_2}\Bigg],
\end{align}
where $\theta_i:=\pi-\theta-\varphi_i$ ($i=1$, 2).
The geometric parameters $f_1$,  $f_2$, and $f_3$ are contributions from the contact line, surface and bulk energy terms, respectively.
The factor $f$ retains explicit dependence on contact angle 
$\theta$ and pore geometry $(\phi_1, \phi_2)$, allowing us to assess quantitative influence of these factors to the nucleation behavior.

\begin{figure}[t!]
\centering 
\includegraphics[width=0.48\textwidth]{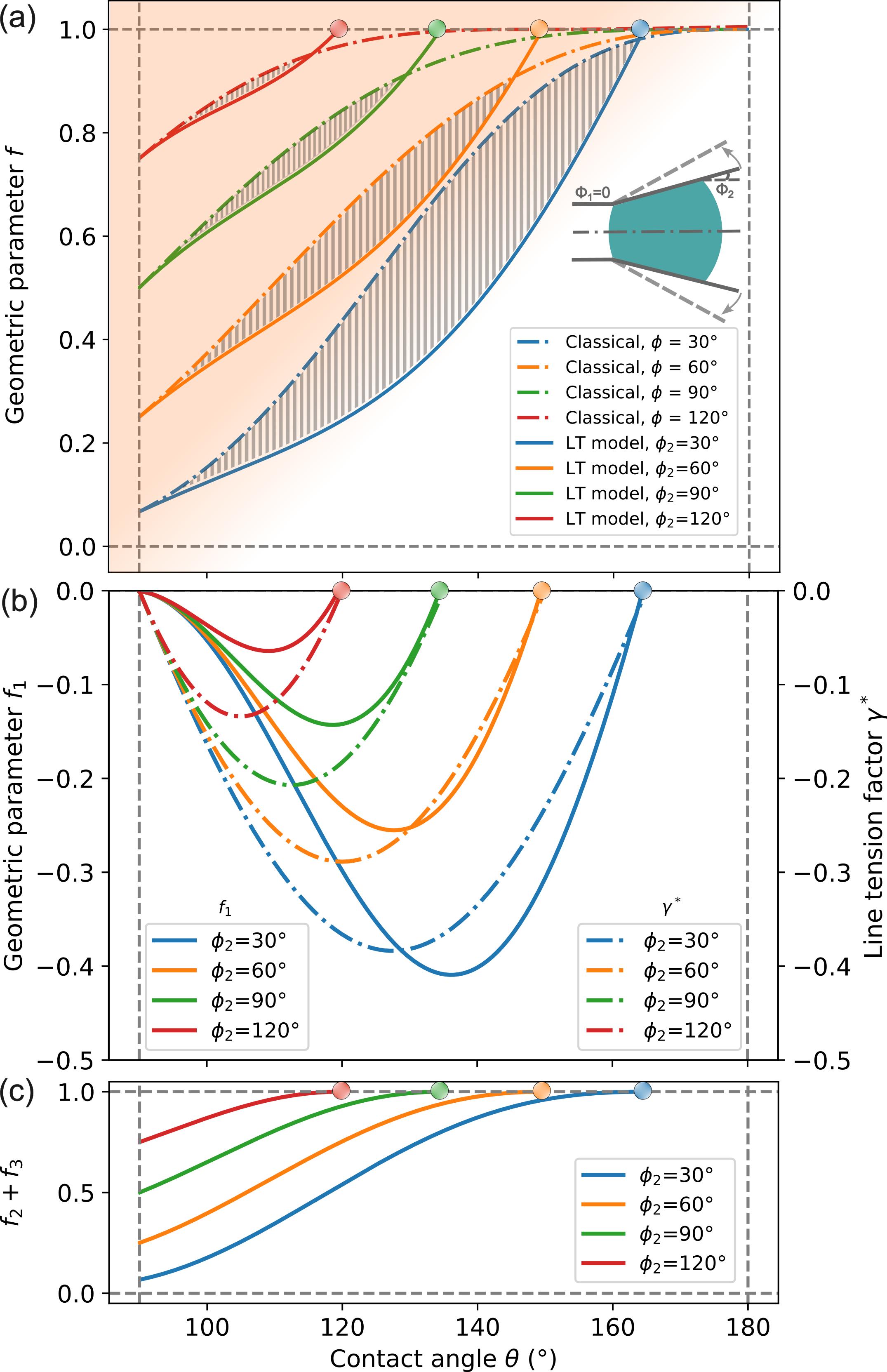}% Here is how to import EPS art
\caption{Geometrical parameters of critical nuclei as a function of contact angles $\theta$ for a fixed left opening angle $2\phi_1=0$ and varying right opening angles $2\phi_2$. (a) Comparison between predictions from the line-tension (LT) model and classical models for droplets  in conical pores without pinning. Shaded regions indicate deviations arising from the inclusion of line tension. The inset illustrates the pore geometry and the nucleated droplet configuration. (b) and (c) show the variations of geometrical parameters from line energy and surface energy contributions, respectively, with respect to $\theta$. Panel (b) also includes the corresponding dimensionless line tension
$\gamma^*$ as a reference. Colored dots in (a)--(c) highlight the conditions where homogeneous nucleation occurs, representing the upper limit of $\theta$ for which the LT model remains valid.
}
\label{fig:f-phi1/2-}
\end{figure}

Remarkably, we find the line tension effect significantly alters nucleation, with the energy barrier highly tunable via pore asymmetry and wettability.
Fig.~\ref{fig:f-phi1/2-}(a) illustrates the geometric parameter 
$f$ as a function of the contact angle for different opening angles of the pore. 
%Geometry-induced deviation from classical nucleation (Fig. 2)
We first plot the predictions from the classical nucleation theory for critical droplets in  conical pores with different opening angles, as indicated by dot dashed lines. For comparison, we also present predictions from our model incorporating line tension effects at the surface edge (referred to as the LT model), shown as solid lines. In this case, we consider asymmetric pores with a fixed left opening angle $2\phi_1=0$ and varying right opening angles $2\phi_2$. Additional results for  $\phi_1 = -20^\circ$, $60^\circ$, and $90^\circ$ are provided in the Supplemental Document.
Our LT model predicts a pronounced suppression of the nucleation barrier compared to the classical model, especially at moderate wettability, namely, from $\theta=90^\circ$ to $\theta=\theta_i$, where $\theta_i$ denotes the intersection point between the solid and dot dashed lines. The shaded region quantifies the suppression of nucleation attributable to line tension effect, which classical theory neglects. In contrast, under strongly hydrophobic conditions with $\theta>\theta_i$, the nucleation barrier in asymmetric pores exceeds that of conical pores. The Inset schematically illustrate a nucleated droplet within an  asymmetric pore geometry characterized by $\phi_1=0$ and an adjustable $\phi_2$.

Fig.~\ref{fig:f-phi1/2-}(b) shows the geometric factor $f_1$ (solid lines) and the dimensionless line tension factor $\gamma^*$ (dot dashed lines) as functions of the contact angle, both predicted by the LT model. It indicates that surface-edge pinning becomes profound at moderate contact angles within the range $\theta \in (90^\circ, \theta_c)$, where the threshold angle $\theta_c=180^\circ-0.5\phi_2$. Within this regime, the line tension contributes a net negative energy. %, stabilizing nuclei formation. 
A striking finding is the non-monotonic behavior of the line tension factor and the associated geometric factor $f_1$ with contact angle. Local minima in  $f_1$ are observed within $\theta \in (90^\circ, \theta_c)$ and these minima correspond to the largest deviations between LT moedels and the classical models.   

Fig.~\ref{fig:f-phi1/2-}(c) shows that the sum of geometric parameters from the volume and surface energy contribution $f_2+f_3$ monotonically increases with contact angle. Since the contribution from line tension is relatively smaller, $f$ still monotonically increases with contact angle, but the increase rate is different from that of $f_2+f_3$. It should be noted that, the contact angle range $\theta \in (90^\circ, \theta_c)$ is divided into two intervals, i.e. I) $\theta \in (90^\circ, \theta_i)$ and II) $\theta \in (\theta_i, \theta_c)$. In the first interval (shaded areas in Fig.~\ref{fig:f-phi1/2-}(a)), the negative line energy leads to the lower nucleation barrier for asymmetric pores, while in the second interval, the pining effect changes the droplet profile, leading to the increas of the sum $f_2+f_3$, which dominant over the line tension effect, so that the nucleation barrier for asymmetric pores becomes higher than that of the conical pores.

When $\theta = \theta_c$, the droplet becomes suspended, symmetrically positioned between $S_1$ and $S_2$, and the pining effect vanishes. In this configuration, the spherical droplet exhibits identical contact angles with both the left and right pore walls (or $S_1$ and $S_2$).  This geometrically induced upper limit for the contact angle ($\theta_c =180^\circ-0.5\phi_2$) represents a critical threshold for heterogeneous nucleation, as indicated by the colored circles in Fig.~\ref{fig:f-phi1/2-}(a)--(c). This limit is absent in classical frameworks for conical pores and establishes a fundamental constraint on nucleation or phase transition in nano-confinement. This state corresponds to the critical case where the droplet becomes spherical and achieves a homogeneous nucleation state with $f=1$.  The upper limit contact angle $\theta_c =180^\circ-0.5\phi_2$ aligns with our previous theoretical prediction for capillary adsorption of droplets into a funnel-like structure confirmed by phase-field simulations~\cite{wu2022capillary}.
For even more hydrophobic scenarios with $\theta>\theta_c$, the droplet maintains a spherical shape, effectively detaching from the right wall of the pore. In such cases, the nucleation barrier becomes identical to that of homogeneous nucleation, again characterized by $f=1$.

As the right opening angle $2\phi_2$ decreases, the line tension contribution becomes more profound, facilitating nucleation over a broader range of contact angles. This trend is evident from the enlarged shaded regions in Fig.~\ref{fig:f-phi1/2-}(a) as $\phi_2$ decreases from $120^\circ$ to $30^\circ$. 
Specifically, when $\phi_2>120^\circ$ (i.e., the pore takes on a thorn-like geometry), the shaded area is expected to largely decrease due to the reduced contribution of line energy and the lowering of $\theta_c$. In this case, the critical nucleus droplet shows a spherical shape over a wider range of contact angles ($\theta >\theta_c=180^\circ-0.5\phi_2$), consistent with the classical Cassie-Baxter model. This model predicts that nano-/microstructures can induce superhydrophobicity by minimizing droplet–substrate contact, even at low intrinsic contact angles~\cite{wang2023wetting}.
%This rich interplay highlights how nanoscale geometry combined with wettability directly modulates nucleation, enabling in situ tuning of phase transitions.
%This extends the classical nucleation framework to a regime where geometry becomes a control parameter.

\begin{figure}
\centering
\includegraphics[width=0.48\textwidth]{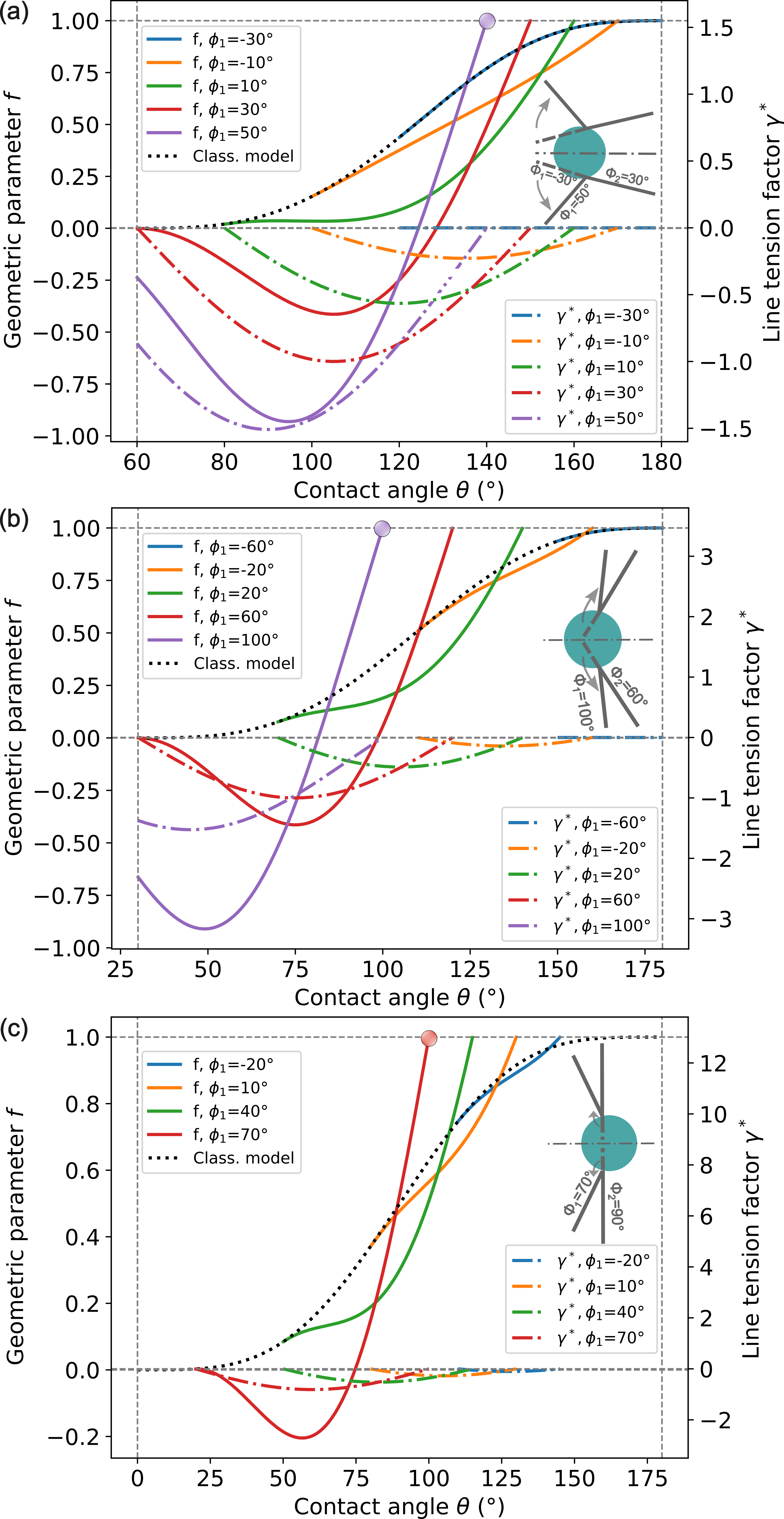}% Here is how to import EPS art
\caption{Geometrical parameters and line tension factors of critical nuclei as functions of the contact angle $\theta$ for fixed $\phi_2$ and varying $\phi_1$. The insets depict the critical nucleus configurations within the pore structures, matching the colored dots. Panels (a)--(c) represent cases with $\phi_2=$ 30$^\circ$, 60$^\circ$, 90$^\circ$, respectively. Predictions from classical models for droplets in smooth conical pores with $\phi=\phi_2$  are indicated as dotted lines for comparison.
}
\label{fig:fig3} 
\end{figure}
To further explore the effect of pore asymmetry, we fix $\phi_2$
and vary $\phi_1$ to examine a broader range of pore geometries. Fig.~\ref{fig:fig3}(a)–(c) illustrate how the geometric parameter $f$ and the line tension factor $\gamma^*$ evolve with  
$\theta$ for increasing $\phi_1$ under fixed values of $\phi_2=$ 30$^\circ$, 60$^\circ$, 90$^\circ$, respectively. 
The valid contact angle range for the LT model is defined as ($\theta_0$, $\theta_c$], where $\theta_0=
\max\left\{90^\circ - \phi_1,\ 90^\circ - \phi_2\right\}$
(see Supplemental Document for an explanation).
As presented in Fig.~\ref{fig:fig3}(a), when $\phi_1=-30^\circ$, the pore shows a 
conical geometry, thus the LT model prediction aligns with that of the classical model with $\phi=$ 30$^\circ$ particularly in the highly hydrophobic regime. This agreement arises because the LT model assumes equal curvature radii on both sides of the droplet. In this special case, the LT model recovers the classical solution corresponding to the case shown in Fig.\ref{fig1:schematic-pore}(b), rather than (a). 
As $\phi_1$ increases to $\phi_1=-10^\circ$ and $10^\circ$, the geometric factor $f$ continues to increase with $\theta$, though at different rates depending on the specific left-side angle $2\phi_1$. A further increase to $\phi_1 = 30^\circ$ and $50^\circ$ introduces a non-monotonic trend in the $f$–$\theta$ relationship, attributed to the enhanced influence of line tension. The occurrence of negative values of $f$ in certain $\theta$ ranges reflects the dominance of line tension in driving nucleation under those geometric configurations.

Similar to the previous discussion in Fig.~\ref{fig:f-phi1/2-}, there is an upper limit contact angle $\theta_c=180^\circ-0.5(\phi_1+\phi_2)$ for the LT model which corresponds to the critical case where the droplet transitions into a homogeneous nucleation state with $f=1$ and $f_1=0$. This critical state is indicated by the violet circle in Fig.~\ref{fig:fig3}(a) for $\phi_1 = 50^\circ$, and by the violet and red circles in Fig.~\ref{fig:fig3}(b,c) for $\phi_1 = 100^\circ$ and $70^\circ$, respectively, with corresponding schematic insets.
It is also found that both $\theta_i$ and $\theta_c$ decrease with increasing 
$\phi_1$.
%contact angle within the range
%$\theta \in (-30^\circ, 50^\circ)$. 
Similar to Fig.~\ref{fig:f-phi1/2-}, the whole contact angle range is divided into two regimes: I) a lower-barrier regime for $\theta < \theta_i$, where line tension reduces the nucleation barrier; II) a higher-barrier regime for $\theta > \theta_i$, where the nucleation barrier increases relative to classical predictions.
Both asymmetric and conical pores can promote nucleation relative to the homogeneous case, reaffirming that homogeneous nucleation represents the upper limit of the nucleation barrier~\cite{liu2000heterogeneous}. 
%Therefore, the nucleation barrier can be actively controlled by manipulating the interplay between $\theta$ and $\phi_1$, as revealed in the energy landscape shown in Fig.~\ref{fig:fig3}(a). 
Fig.~\ref{fig:fig3}(b) and (c) present analogous trends for larger right-side angles $\phi_2 = 60^\circ$ and $90^\circ$, respectively. However, the tunable range of contact angles changes as $\phi_2$ varies and the influence of line tension significantly changes compared to the case with a smaller half-opening angle ($\phi_2 = 30^\circ$).
Overall, the LT model enables the construction of a comprehensive energy landscape for nucleation in diverse pore geometries, including asymmetric cases, conical shapes, stepped edges, and crevices, highlighting the model’s broad applicability to realistic nano-/microstructured systems.
%This provides a powerful framework for precisely controlling nucleation processes through tailored geometries and wettability.

The above detailed investigation reveals one nontrivial feature. For a given pore geometry, the line tension term becomes especially significant when a particular contact angle angle is achieved. 
For example, with $\phi_1 = 0^\circ$ and $\phi_2 = 30^\circ$ (see Fig.~\ref{fig:f-phi1/2-}), the line tension factor reaches a minimum $\gamma^*_{\text{min}} \approx -0.4$ at $\theta=125^\circ$. Assuming a critical radius of  water nucleus $R_c=50$ nm~\cite{sear2007nucleation} and a surface tension $\sigma_{lg} = 0.07$ N/m, Eq.~\eqref{eq:critical-linetension} yields a line tension of approximately $-1.4 \times 10^{-9}$ N—an order of magnitude larger than previously reported values of $10^{-11}$ to $10^{-10}$ N for water droplets on the flat quartz considering intermolecular forces~\cite{pompe2000three}. By tuning $R_c$, pore geometry, and wettability, the line tension can even span several orders of magnitude, helping to explain the wide variability seen in the literature~\cite{whyman2008rigorous}.
This may also explain the large deviations from classical predictions observed in nanoscale systems such as the anomalous stability of nanobubbles in cylindrical pores~\cite{guillemot2012activated} or on surfaces under pining~\cite{tan2018surface}, pore size and shape dependence of nucleation rate~\cite{chayen2006experiment,page2006heterogeneous,sear2007nucleation,diao2011role},  shifts in ice nucleation due to nanoscale texture~\cite{gurganus2014nucleation}, and wettability-dependent evaporation rates in nanopores~\cite{fan2020evaporation}. 
To conclude, we extend classical nucleation theory to explicitly incorporate line tension arising from surface-edge pinning, uncovering a confinement-dominated regime of nucleation. 
Unlike classical line-tension models for smooth surfaces that often assume an ill-defined molecular-scale force, our approach treats line tension as a mechanical consequence of curvature and Laplace pressure for droplets confined near surface edges.
It not only provides a quantitative expression for the edge-pinned line tension in confined systems, but also exposes how geometry and wetting jointly govern nucleation thresholds. These insights extend far beyond idealized nucleation models and offer a physically grounded tool for understanding and controlling nucleation in confined systems with broad applications in areas such as targeted drug delivery, pharmaceutical crystallization, and nanofabrication.
The geometry-tailored line tension effect may provide a route to control nucleation at will, replacing trial-and-error design with predictive strategies. While direct experimental validation is beyond this work’s scope, the proposed model is testable via precision experiments on tailored pore geometries.
Looking forward, this framework can be generalized to incorporate fluctuations~\cite{fitzner2017pre}, 
instabilities~\cite{theisen2007capillary}, 
active wetting~\cite{liese2025chemically}, and even two-step nucleation~\cite{erdemir2009nucleation},
%and dynamic evolution of nucleated droplets or bubbles~\cite{gallo2021heterogeneous}, 
laying the foundation for a unified theory of phase transitions ~\cite{davoodabadi2021evaporation} and nanobubbles~\cite{lohse2015surface} under nano-scale confinement, where geometrical edge effects are essential.

%Looking ahead, this framework can be extended beyond static energy landscapes to  thereby offering a more comprehensive understanding of confined phase transitions. More broadly, this framework lays the foundation for a generalized theory of confined phase behavior, extending beyond droplet nucleation to crystallization, phase separation, and biological self-assembly~\cite{deshpande2019spatiotemporal,yuan2019nucleation,meldrum2020crystallization},  

\begin{acknowledgments}
This research was supported by the U.S. Food and Drug Administration under the FDA BAA-22-00123 program, Award Number 75F40122C00200.
\end{acknowledgments}

%\nocite{*}

\bibliography{apssamp}% Produces the bibliography via BibTeX.

\end{document}